\documentclass[12pt]{article}
\begin{document}


\input{psfig}


\def \ms {{\overline{\mbox{MS}}}}
\newcommand{\z}{&&\hspace*{-0.7cm}}
\newcommand{\zz}{&&\hspace*{-0.3cm}}
\newcommand{\bea}{\begin{eqnarray}}
\newcommand{\eea}{\end{eqnarray}}
\newcommand{\prepr}[1] {\begin{flushright} {\bf #1} \end{flushright} 
\vskip 1.5cm}
%



\begin{center}
{\bfseries
SMALL $X$ BEHAVIOR OF PARTON DISTRIBUTIONS WITH FLAT INITIAL CONDITIONS.\\
A STUDY OF HIGHER-TWIST EFFECTS}

\vskip 5mm

A.V. Kotikov$^{1 \dag}$ and  G. Parente$^{2 \dag} $ 

\vskip 5mm

{\small
(1) {\it
Institut fuer Theoretical Teilehenphysik, Universitaet Karsruhe,
D-76128 Karsruhe, Germany and 
BLThPh, Joint Institute for Nuclear Research,~
141980 Dubna, Russia
}
\\
(2) {\it
Departamento 
de F\'\i sica de Part\'\i culas, 
Universidade 
de Santiago de Compostela,
\\
15706 Santiago de Compostela, Spain
}
\\
$\dag$ {\it
E-mail: kotikov@thsun1.jinr.ru, gonzalo@fpaxp1.usc.es
}}
\end{center}

\vskip 5mm

\begin{center}
\begin{minipage}{150mm}
\centerline{\bf Abstract}

We study 
 the  $Q^2$ evolution of parton distributions
at small $x$ values,
obtained in the case
of flat initial conditions.
The contributions of twist-two and (renormalon-type) higher-twist 
operators of the Wilson operator product expansion are taken into account.
The results are in good
agreement with deep inelastic scattering
experimental data from HERA.
\end{minipage}
\end{center}

\vskip 10mm

\section{Introduction}

The measurements of the deep-inelastic scattering
structure function (SF)
$F_2$ in HERA
\cite{H1,H1n}
have permitted the access to
a very interesting kinematical range for testing the theoretical
ideas on the behavior of quarks and gluons carrying a very low fraction
of momentum of the proton, the so-called small $x$ region.
In this limit one expects that
non-perturbative effects may give essential contributions. However, the
reasonable agreement between HERA data and the 
next-to-leading order (NLO)
approximation of
perturbative
QCD that has been observed for $Q^2 > 1 $GeV$^2$ (see the  review
in Ref. \cite{CoDeRo}) indicates that
perturbative QCD could describe the SF
evolution 
up to very low $Q^2$ values,
traditionally explained by soft processes
\footnote{The agreement can be explained \cite{DoShi} by an effective
coupling constant scale essentially higher then $Q^2$ (see review
\cite{Andersson} and references therein).}.
It is of fundamental importance to find out the kinematical region where
the well-established perturbative QCD formalism
can be safely applied at small $x$.

The standard program to study the small $x$ behavior of
quarks and gluons
is carried out by comparison of data
with the numerical solution of the
DGLAP
equations 
by fitting the parameters of the
$x$ profile of partons at some initial $Q_0^2$ and
the QCD energy scale $\Lambda$ (see, for example, \cite{MRS,KKPS}).
However, if one is interested in analyzing exclusively the
small $x$ region ($x \leq 0.01$), 
there is the alternative of doing a simpler analysis
by using some of the existing analytical solutions of DGLAP 
in the small $x$ limit (see, for example, Ref. \cite{CoDeRo} for review).
This was done so in Refs. \cite{BF1}-\cite{Q2evo3}
where it was pointed out that the HERA small $x$ data can be
interpreted in 
terms of the so called doubled asymptotic scaling phenomenon
related to the asymptotic 
behavior of the DGLAP evolution 
discovered  in \cite{Rujula}
many years ago. 

Here we compile
results obtained 
in \cite{Q2evo}-\cite{HT},
where the contributions of higher-twist (HT) operators 
(i.e. twist-four ones and twist-six ones)
of
the Wilson operator product expansion are taken into account
\footnote{Refs.\cite{Q2evo1,HT} and the present note
deal with 
the contributions of HT
operators only in the renormalon model approach (see \cite{SMaMaS}).
The complete analysis of the HT contributions to the SF $F_2$, 
$dF_2/d\ln (Q^2)$ and $F_L$ will be presented in \cite{Q2evo3}.}.
The importance of the contributions of HT
operators at small-$x$
has been demonstrated in many studies (see \cite{Bartels}).

We would like to note that
the results of \cite{Q2evo}
are the extension to the NLO QCD approximation of previous 
leading order (LO)
studies \cite{Rujula,BF1}.
The main ingredients are:

{\bf 1.} Both, the gluon and quark singlet densities are
presented in terms of two components ($'+'$ and $'-'$)
which are obtained from the analytical $Q^2$
dependent expressions of the corresponding ($'+'$ and $'-'$)
parton distributions moments.

{\bf 2.} The $'-'$ component is constant
at small $x$, whereas the 
$'+'$ component grows at $Q^2 \geq Q^2_0$ as 
$\sim \exp{(\sigma_{NLO})}$, where
\bea
\sigma_{NLO} = 2\sqrt{(\hat d_{+}s+\hat D_{+}p)\ln x},
\nonumber
\eea
and the LO term $\hat d_+ = -12/\beta_0$ and the NLO one 
$\hat D_{+}=\hat d_{++}+\hat d_{+}\beta_1/\beta_0$ with
$\hat d_{++} = 412f/(27\beta_0)$. 
Here the coupling constant
$a_s=\alpha_s/(4\pi)$, 
$s=\ln (a_s (Q^2_0)/a_s (Q^2))$ and
$p=a_s (Q^2_0) - a_s (Q^2)$,
$\beta_0$ and $\beta_1$ are the first two 
coefficients of QCD 
$\beta$-function and $f$ is the number of active flavors.

\section{Basical formulae
}

Thus, our purpose
is to demonstrate the small $x$ asymptotic
form of parton distributions
in the framework of the DGLAP equation starting at some $Q^2_0$ with
the flat function:
 \begin{eqnarray}
f^{\tau2}_a (Q^2_0) ~=~
A_a ~~~~(\mbox{ hereafter } a=q,g), \label{1}
 \end{eqnarray}
where $f^{\tau2}_a$ are the leading-twist (LT) parts of
parton (quark and gluon)
distributions multiplied by $x$
and $A_a$ are unknown parameters that have to be determined from data.
Through this work at small $x$ we neglect
the non-singlet quark component.

We would like to note that new HERA data \cite{H1n} demonstrate a rise
of SF $F_2$ 
at low $Q^2$ values ($Q^2 < 1 $GeV$^2$)
when $x \to 0$ (see Fig.2, for example). The rise can be explained
in natural way by incorporation  of HT 
terms in our analysis (see section 2.2).

We shortly compile below the main results found in \cite{Q2evo,Q2evo3} 
at the LO
approximation (the LT
results at the NLO approximation
may be found in \cite{Q2evo}). 
The full small $x$ asymptotic results
for parton distributions and SF $F_2$ 
at LO of perturbation theory is:
 \begin{eqnarray}
F_2(x,Q^2)~=~ e \cdot \left[
f^{\tau2}_q(x,Q^2) + f^{h\tau}_q(x,Q^2) +
 f^{h\tau}_g(x,Q^2) \right],
\label{r10} 
\end{eqnarray}
where $e= (\sum^f_1e^2_i)/f$ is the average charge square.
The LT
parts $f^{\tau2}_a(x,Q^2)$
and the HT
ones $f^{h\tau}_a(x,Q^2)$ can be
represented as sums of the $'+'$ and $'-'$ components 
 \begin{eqnarray}
 f^{\tau2}_a(x,Q^2) &=& f^{\tau2,+}_a(x,Q^2) + 
f^{\tau2,-}_a(x,Q^2) \; , 
\label{r11} \nonumber \\
 f^{h\tau}_a(x,Q^2) &=& f^{h\tau,+}_a(x,Q^2) + 
f^{h\tau,-}_a(x,Q^2). \;  
\label{r12}
\end{eqnarray}


\subsection{The LT
contribution} \indent

The small $x$ asymptotic results for parton distributions have the form
\begin{eqnarray}
f^{\tau2,+}_g(x,Q^2)&=& \biggl(A_g + \frac{4}{9} A_q \biggl)
\tilde I_0(\sigma) \; e^{-\overline d_{+}(1) s} ~+~O(\rho) 
~~\;\; ,\label{8.0} \\
f^{\tau2,+}_q(x,Q^2)&=& \frac{f}{9}\biggl(A_g + \frac{4}{9} A_q \biggl) 
\rho \; \tilde I_1(\sigma) \;
e^{-\overline d_{+}(1) s} ~+~O(\rho) \; , \label{8.01} 
\\
f^{\tau2,-}_g(x,Q^2)&=& - \frac{4}{9} A_q e^{- d_{-}(1) s} 
~+~O(x) ,
\label{8.00} \\
f^{\tau2,-}_q(x,Q^2)&=&  A_q e^{- d_{-}(1) s} ~+~O(x) \; ,\label{8.02}
 \end{eqnarray}
where
$\overline d_{+}(1) = 1+20f/(27\beta_0)$ and
$          d_{-}(1) = 16f/(27\beta_0)$
are the regular parts of $d_{+}$ and $d_{-}$
anomalous dimensions, respectively, in the limit $n\to1$ 
\footnote{From now on, for a quantity $k(n)$ we use the notation
$\hat k(n)$ for the singular part when $n\to1$ and
$\overline k(n)$ for the corresponding regular part. }. 
The functions $\tilde I_{\nu}$ ($\nu=0,1$) 
are related to the modified Bessel
function $I_{\nu}$
and to the Bessel function $J_{\nu}$ by:
\begin{eqnarray}
\rho^{\nu}\tilde I_{\nu}(\sigma) ~=~
\left\{
\begin{array}{ll}
~~~~~\rho^{\nu} I_{\nu}(\sigma), & \mbox{ if } s \geq 0 \\
{(-\overline \rho)}^{\nu}J_{\nu}(\overline \sigma),~ 
\overline \rho = i \rho,~ \overline \sigma = i \sigma, 
& \mbox{ if } s  <   0
\end{array} \right.  
.
\label{4}
\end{eqnarray}
The 
variables $\sigma$ and $\rho$ are
given by
\begin{eqnarray}
\sigma =2\sqrt{\hat d_{+} s \,\ln(x)} \; , ~~~
\rho = \sqrt{\frac{\hat d_{+} s}{\ln(x)}}
= \frac{\sigma}{2\ln(1/x)}
\label{slo}
\end{eqnarray}


\subsection{The HT
contribution} \indent


Using the results in \cite{Q2evo1,HT} (which are based on 
calculations \cite{SMaMaS,method}), we demonstrate the effect
of HT
corrections in the renormalon model case
(see recent review of renormalon models in \cite{Beneke}).
We sketch below the basical results 
making the following substitutions
in the corresponding twist-two results presented in
Eqs.(\ref{8.0})-(\ref{8.02}) (complete formulae can be found in 
\cite{Q2evo3}):

$f^{\tau2,+}_g(x,Q^2)$ and $f^{\tau2,+}_q(x,Q^2)$
(see Eqs.(\ref{8.0}) and (\ref{8.01})) 
$\to f^{h\tau,+}_g(x,Q^2)$ and $f^{h\tau,+}_q(x,Q^2)$ 
by
\bea 
A_a \tilde I_0(\sigma) &\to & A_a \, \,
\frac{16f}{15\beta_0^2}  
\sum_{m=1}^2 k_m
\frac{(\Lambda_{m,a})^{2m}}{Q^2} \left(
\frac{2}{\rho} \tilde I_1(\sigma)
-\ln \left(
\frac{\Lambda^2_{m,a}}{Q^2} \right) 
\tilde I_0(\sigma)  \right),  
\label{r01} \\
A_a \rho \tilde I_1(\sigma) 
&\to & A_a \, \,
\frac{128f}{45\beta_0^2} \sum_{m=1}^2 k_m
\frac{(\Lambda_{m,a})^{2m}}{Q^2} \left(
\frac{2}{\rho} \tilde I_1(\sigma)
-\ln \left(
\frac{\Lambda^2_{m,a}}{Q^2} \right) 
\tilde I_0(\sigma)  \right),
\label{r1}
\eea
where $k_1=1$, $k_2=-8/7$ and
$\Lambda_{1,a}$ and $\Lambda_{2,a}$ are magnitudes of twist-four and 
twist-six corrections. 

$f^{\tau2,-}_g(x,Q^2)$ and $f^{\tau2,-}_q(x,Q^2)$ (see Eqs.(\ref{8.00})
and (\ref{8.02}))
 $ \to f^{h\tau,-}_g(x,Q^2)$ and $ f^{h\tau,-}_q(x,Q^2)$
by
\bea 
A_q &\to &  A_q \, \,
\frac{16f}{15\beta_0^2} 
\sum_{m=1}^2 k_m 
\frac{(\Lambda_{m,q})^{2m}}{Q^2}
\ln \left(\frac{Q^2}{x^2\Lambda^2_{m,q}} \right),
\label{r6} \\
A_q &\to &  A_q \, \,
\frac{128f}{45\beta_0^2} 
\sum_{m=1}^2 k_m 
\frac{(\Lambda_{m,q})^{2m}}{Q^2}
\left( 
\ln \left(\frac{Q^2}{x\Lambda^2_{m,q}} \right) -\frac{11}{3} \right)
\ln \left(\frac{1}{x} \right).
\label{r7}
\eea

\begin{figure}[t]
\vskip -0.5cm
\psfig{figure=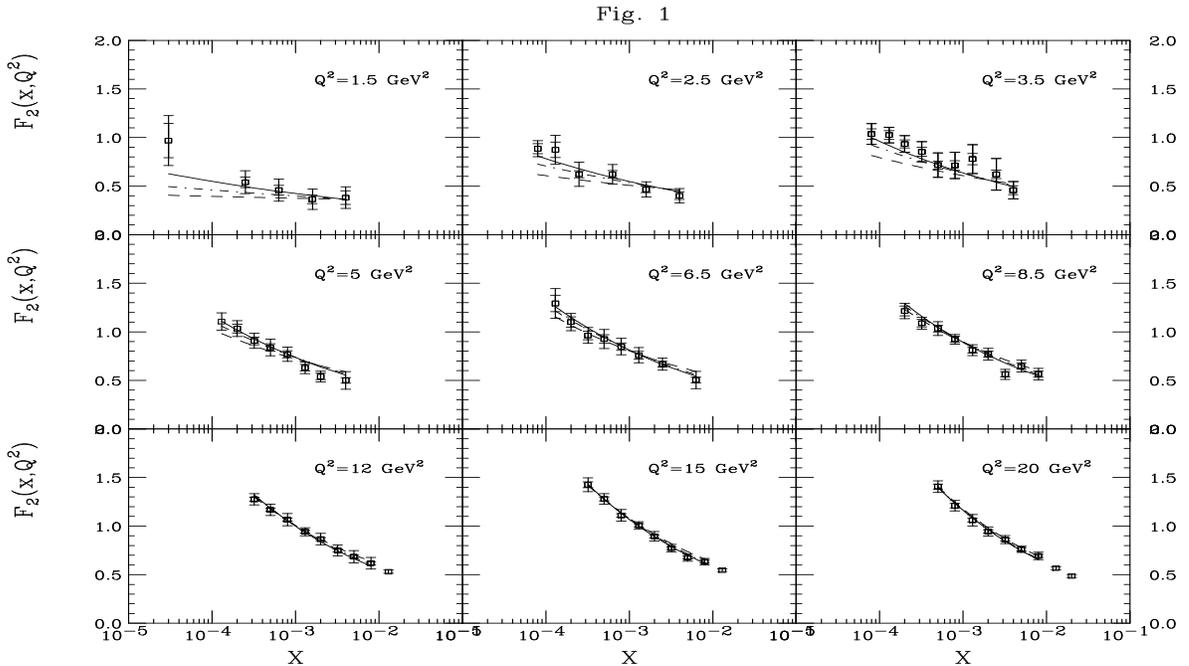,height=3.5in,width=6.2in}
%
\caption{The SF
$F_2$ as a function of $x$ for different
$Q^2$ bins. The experimental points are from H1 \cite{H1}. 
The inner error 
bars are statistic while the outer bars represent statistic and systematic 
errors added in quadrature. The dashed and dot-dashed curves are obtained 
from fits (based on LT
formulae)
at LO and NLO respectively with fixed $Q^2_0=1$ GeV$^2$. The solid
line is from the fit at NLO giving $Q^2_0=0.55$ GeV$^2$.
}
\end{figure}

From Eqs.(\ref{r01})-(\ref{r7}) one can notice that the higher-twist
terms modify the flat condition Eq.(\ref{1}). They lead to a rise of
parton distributions and, thus, SF $F_2$ 
at low value $Q^2_0$, when $x \to 0$. This is in excellent agreement
with the recent low $Q^2$ HERA data \cite{H1n}.


\section{Results of the fits}

With the help of the results obtained in the previous section we have
analyzed $F_2$ HERA data at small $x$ from the H1 and ZEUS 
collaborations
\cite{H1,H1n}.
In order to keep the analysis as simple as possible
we have fixed the number of active flavors $f$=4 and
$\Lambda_{\ms}(n_f=4) = 250$ MeV, which
is a reasonable value extracted from the traditional (higher $x$)
experiments.
Moreover, we put $\Lambda_{1,a}=\Lambda_{2,a}$ in agreement with
\cite{DaWe}.
The initial scale of the 
parton densities was also fixed
into the fits to $Q^2_0 = 1$ GeV$^2$, although later it was released
to study the sensitivity of the fit to the variation of this parameter.
The analyzed data region was restricted to $x<0.01$ to remain within the
kinematical range where our results are
accurate.

Fig. 1 shows $F_2$ calculated from the fit (based only on LT
formulae) with Q$^2$ $>$ 1 GeV$^2$
in comparison with 1994 H1 data
(first article in \cite{H1}).
Only the lower $Q^2$ bins are shown.
One can observe that the NLO result (dot-dashed line)
lies closer to the data
than the LO curve (dashed line).
The lack of agreement between data and lines observed
at the lowest $x$ and $Q^2$ bins suggests
that the flat behavior should occur at $Q^2$ lower
than 1 GeV$^2$.
In order to study this point we have done the
analysis considering $Q_0^2$ as a free parameter.
Comparing the results of the fits (see \cite{Q2evo})
one can notice
the better agreement with the experiment 
at fitted $Q^2_0=0.55$ GeV$^2$ (solid curve)
is apparent at the lowest kinematical bins.

\begin{figure}[t]
\vskip -0.5cm
\psfig{figure=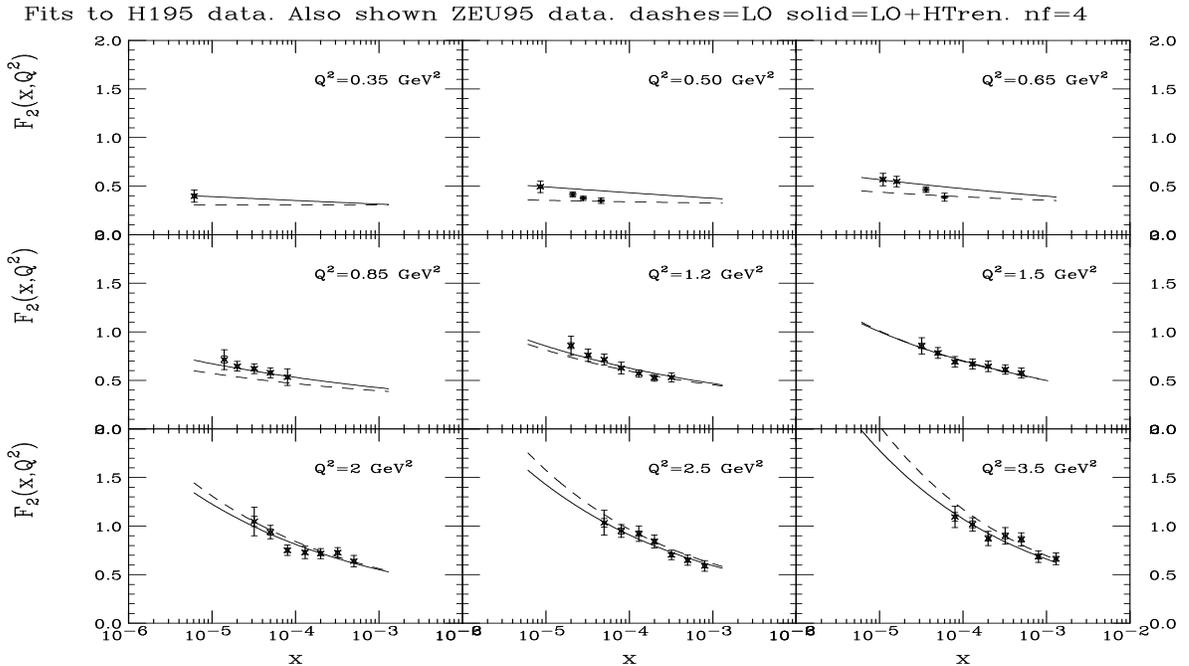,height=3.5in,width=6.2in}
%
\caption{The SF
$F_2$ as a function of $x$ for different
$Q^2$ bins. The experimental points are from H1 and ZEUS \cite{H1n}. 
The inner error 
bars are statistic while the outer bars represent statistic and systematic 
errors added in quadrature. The solid curves are obtained 
from fits at LO, when  contributions of HT
terms have been incorporated. 
The dashed curves show only twist-two contributions.
}
\end{figure}

Fig. 2 shows $F_2$ calculated from the fit at LO (based on LT $\&$ HT
formulae) in comparison with 1995 H1 and ZEUS data
\cite{H1n}.
One can observe that these results 
(solid line)
lies closer to the data
than the twist-two results (dashed line).
We have done the
analysis considering $Q_0^2$ as a free parameter.
Comparing the results of the fits (see \cite{Q2evo3})
one can notice
the better agreement with the experiment 
at fitted $Q^2_0=0.61$ GeV$^2$, 
which is close to $Q_0^2$ in the
analysis of 1994 H1 data (see Fig. 1).

\section{Conclusions} 

We have shown that the results developed recently in \cite{Q2evo}-
\cite{Q2evo3}
 have relatively quite simple form and reproduce many
properties of parton distributions at small $x$,
that have been known from global fits.

We found very good agreement between our approach based on QCD 
and HERA data, as it has been observed earlier with
other approaches (see the review \cite{CoDeRo}). 
The (renormalon-type) higher-twist terms lead to the natural explanation of
the rise of $F_2$ at low $x$ for the lowest values of $Q^2$ 
($\leq 1$ GeV$^2$).
The rise has been discovered in recent HERA experiments \cite{H1n}.

As next step of our investigations, we plan to study contributions of
HT operators to relations between parton distributions and
deep inelastic structure functions, observed, for example, in 
\cite{KoPa,KOPAFL}.\\

  {\it Acknowledgments.}
One of the authors (A.V.K.)
 would like to express his sincerely thanks to the Organizing
Committee of XVI International Baldin Seminar on High Energy Physics 
Problems ``Relativistic Nuclear Physics and Quantum Chromodynamics'' 
for the kind invitation. 
A.V.K. and G.P. 
were supported in part, respectively, by 
INTAS  grant N366 (at the first time of the study) and by
Alexander von Humboldt fellowship (at the end of the study)
and by Galician research funds
(PGIDT00 PX20615PR) and Spanish CICYT (FPA2002-01161).

\end{document}